\documentclass[`]{aa}
\usepackage{natbib}
\usepackage{graphicx}
\usepackage{txfonts}
\def\HH{\mbox{H$_2$}}

\def\nH2{{\rm n}({\rm H}_2)}
\def\NH2{{\rm N}({\rm H}_2)}
\def\pccc{~{\rm cm}^{-3}}

\def\Tstar#1 {\mbox{${\rm T}_{\rm #1}^*$}}
\def\Tsub#1 {\mbox{$T_{\rm #1}$}}
\def\TK  {\Tsub K }

 \def\arcmin{\mbox{$^{\prime}$}}

\def\degr{\mbox{$^{\rm o}$}}

\def\p{\mbox{$^+$}}
\def\cotw {\mbox{$^{12}$CO}}

\def\h13cop{\mbox{{H$^{13}$CO\p}}}

\def\c3h2{\mbox{C$_3$H$_2$}}
 
 \def\R0{R$_0$}

\def\ddeg{{}^\circ\kern-.1em}  
\def\Msun{\mbox{M$_{\rm sun}$}}  

\def\kms{\mbox{km\,s$^{-1}$}}

\def\E#1 {$10^{#1}$}
\def\E#1 {E{#1}}
\def\P#1,{$\nH2\TK~=~#1\times~10^4\pccc$~K}
\def\ec#1,#2,#3,{#1\,(#2)\E{#3}}

\def\H3{\mbox{H$_3$}}

\def\RH2{\mbox {R$_{\rm G}$}}
\def\fH2{\mbox {f$_{\HH}$}}
\def\FH2{\mbox {F$_{\HH}$}}

\title{Multiplicity of nuclear dust lanes and dust lane shocks
  in the Milky Way bar}
\author{H. S. Liszt\inst{} }
\institute{National Radio Astronomy Observatory,
           520 Edgemont Road,
           Charlottesville, VA,
           USA 22903-2475}

\begin{document}
\date{received \today}
\offprints{H. S. Liszt}
\mail{hliszt@nrao.edu}
%
% \abstract{}{}{}{}{} 
% 5 {} token are mandatory

\abstract
  % context heading (optional) leave it empty if necessary  
   {}
  % aims heading (mandatory)
    {We show the existence of a small family of inner-galaxy dust lanes 
 and dust lane standing shocks beyond the two major ones  that were 
  previously known to exist}
  % methods heading (mandatory)
   {We analyze images of CO emission in the inner regions of the Galaxy}
  %results heading (mandatory)
  {The peculiar kinematics of the major dust lane features 
   are repeated in several other distinct instances at $l > 0\degr$, 
   in one case at a  contrary location 100 pc above the galactic 
  equator at l $> 3\degr$ at the upper extremity of Clump 2.  
  Like the previously-known dust lanes, these new
 examples are also associated with localized, exceptionally broad
  line profiles believed to be characteristic of the shredding of 
  neutral gas at the standing dust lane shocks. }
 %conclusions headiing (optional), leave it empty if necessary
 {There may be secondary  dust lane and standing
 shocks in the Milky Way bulge. The vertical structure provides
  a temporal sequence for understanding the secular evolution
  of gas flow in the bar.}

\maketitle

\section {Introduction.}

Constructing a coherent picture of the inner-galaxy gas and its kinematics from
our vantage point within the disk has proved challenging.  In the earliest
days the gas was first observed in H I and decomposed into ``features''
\citep{VdK70,CohDav76,Oort77} then subsequently observed in molecular lines 
 (Fig. 1) revealing somewhat different, often seemingly disparate  behavior. 
The various views were compared and reconciled \citep{BurLis83}, largely 
through realization of
the differences in the conditions of observing of various gas constituents,
and the ``features'' were associated with each other and then with an
underlying physical entity, namely a galactic bar.  The process of 
observing,  comparing and assimilating  other constituents of the bar  is
ongoing, for instance with derivation of the distribution of extinction
and dust from the bar dust lanes \citep{MarFux+08}  and study of
the chemistry based on a dynamical separation of consituents
\citep{RodCom+06}.
 
%1
\begin{figure}
\includegraphics[height=9.8cm]{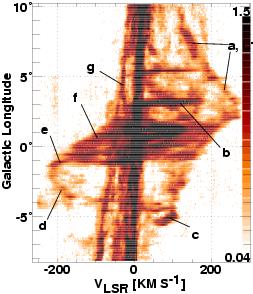}
\caption[]{Longitude-velocity diagram of \cotw\ emission from the dataset
of \cite{BitAlv+97} on a 0.125\degr\ grid, separately averaged over 
$-1\degr \leq b \leq 1\degr$ at each longitude.  Features a 
($aka$ the ``connecting arm'') 
and d are associated with prominent galactic dust lanes; b and c are 
Clump 2 and Clump 1 of \cite{Ban77}, e is the ``rotating nuclear disk'' 
of Rougour and Oort \citep{Oort77}, f is the ``expanding molecular ring'' 
of \cite{KaiKat+72} and g arises in  the 3-kpc arm \cite{Ban80} }.
\end{figure}

Along the way there were many false starts but the notion (which took
hold very early) that the observations must be reconciled with the 
existence of a galactic bar in the presence of an overall tilt 
 of the gas layer \citep{CohDav76,BurLis78} certainly proved correct.  
Fig. 1, a scan along the inner galactic plane in CO, integrated over 
-1\degr $\le b \le $1\degr\ to suppress the wealth of vertical structure, 
illustrates the complexities which must be overcome in order to 
gain a view of the underlying physical phenomena.  It also helps to explain
why our understanding of the observations has improved as models of
the bar gas flow have matured, from simple recognition of the x$_2$-x$_1$ orbit
separation in fast bars \citep{BinGer+91} to holistic views of the hydrodynamic
gas flow over an entire bar \citep{Fux99,RegShe+99,RegTeu03}.  For recent
views of the galactic center gas distribution which incorporate notions
of bar gas flow, see \cite{SawHas+04,Lis06Shred,RodCom+06}
 and \cite{MarFux+08}.

It is access to the rather extravagant vertical structure of the 
inner-galaxy gas which is the truly unique aspect of observing the 
galactic nucleus from the vantage point of the Sun, and this should 
 feed back into models of the gas flow.  However, the vertical 
structure has proved largely intractable theoretically and 
has often been suppressed (as in Fig. 1), ignored or denied, even in 
purely observational descriptions.  Yet, it is the vertical structure 
of the gas which often presents the sharpest and most surprising 
structure \citep{Lis06Shred}.  \cite{MarFux+08} demonstrated a 
vertical separation between the tilts of the dust and molecular gas 
in the Milky Way dust lane which would be important in interpreting 
observations of other systems.

Here we note that the peculiar kinematics which distinguish the 
standing shocks of the large-scale inner-galaxy bar dust lanes 
are repeated several times over the inner regions, in one case with
a tilt quite opposite to that which is usually recognized.  Apparently,
the Milky Way has secondary bar/dust lane structures, as
discussed in Sect. 2. 

\section{Kinematic features associated with the dust lanes and 
dust lane shocks}

The dataset employed here is that of \cite{BitAlv+97}, with \cotw\
spectra on a 0.125\degr\ grid at $|b| \le 1$\degr, $|l| \le 13$\degr\
and on a 0.25\degr\ grid at 1\degr\ $ \le | b | \le$ 2\degr.  
To produce Fig. 1 we averaged (unweighted) over all latitudes at each 
longitude; this figure is similar to one panel of Fig. 3 of \cite{MarFux+08}.
Several of the most prominent kinematic features arising exterior to the
Sgr A-E source complex are called out in Fig. 1.

\subsection{Kinematics of dust line features}
  
Despite their obvious asymmetries in intensity, velocity, and longitude
extent, it is the paired features a and d which traditionally have been
associated with each other and with the {near and far-side} galactic 
dust lanes, 
 respectively \citep{CohDav76,LisBur80,Fux99,RodCom+06,MarFux+08}.  
Features a 
and d are oppositely displaced below and above the galactic equator, 
respectively, in the sense of the most pervasive gas tilt 
\citep{LisBur80}.

The a-feature was sufficiently prominent in H I to have been known
originally as the ``connecting arm'' while the d-feature was not
 always remarked as a separate entity (see Fig. 1 of \cite{CohDav76}).
It is a somewhat of an accident of the Sun's location that the dust 
lane shock appears in such a way as to form the terminal velocity.
Incorporating the contrary kinematics of the a and d-features into the 
derivation of a galactic rotation curve directly from the observed 
terminal velocity, assuming pure circular motion, causes an apparent
decline in equilibrium circular velocity with galactocentric radius,
corrupting the mass model \citep{OstCal81,BurLis93}.

 Feature f (the expanding molecular ring) and its 
positive-velocity counterpart crossing zero-longitude at 
+165 \kms\ arise in the spray of material flowing inward along the 
dust lanes where they contact the nuclear star-forming ring
containing the Sgr A-E radiocontinuum sources,
see \cite{RegShe+99}, Fig. 9, or \cite{Fux99}, Fig. 17.
Sgr B2, which has been described as resulting from a cloud-cloud
collision \citep{HasAra+08}, is probably at such a contact point.

%2
\begin{figure*}
\includegraphics[height=9.45cm]{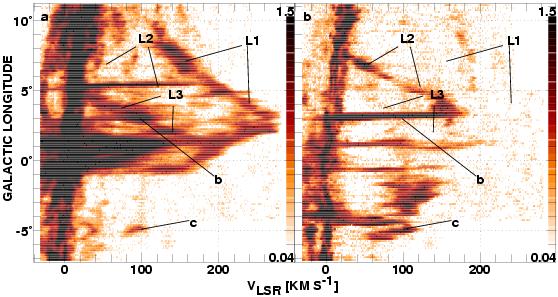}
\caption[]{ Positive-velocity portions of l-v diagrams.  
Left: at negative latitudes averaged over $-1\degr \le b \le 0\degr$.  
Right: at  positive latitudes, averaged over 
$0.5 \degr \le b \le 1\degr$.  The b and c notation 
is repeated from Fig. 1.} 
\end{figure*}

Several other prominent features in Fig. 1 are directly related to 
the dust lanes, in particular the b-feature, Clump 2 of 
\citep{Ban77,StaBan86}, and the broad, localized, anonymous feature at 
l = 5.5\degr.  These were mapped recently in several molecules at 1\arcmin\
resolution \citep{Lis06Shred} and discussed in terms of the shredding of
large quantities of molecular gas at the standing dust lane shock 
\citep{Fux99}:  various estimates yield some $10^6$ \Msun\ in either
feature.  The broad lines in Fig. 1 are spatially resolved at arcminute 
(2.4 pc) resolution, with velocity gradients stronger than seen 
across the circumnuclear disk 
f around SgrA$^\ast$. 

The weakness of the  far-side dust lane feature d at negative 
longitude may be related to the absence of similar broad-lined 
features at negative  longitude, indicating that the negative 
longitude dust lane is (temporarily?) somewhat starved of gas.

\subsection{Other dust lane features}

Fig. 2 shows the higher-velocity portions of two composite l-v
diagrams constructed  below and above the galactic equator, 
 at left and right, respectively.  Like Fig. 1 these have 
also been integrated in latitude, but over smaller regions.  
At negative latitudes in the left-hand panel, a locus labelled L3, 
nearly parallel to the connecting arm feature, is prominent; it is
of course faintly visible in Fig. 1 as well  and was labelled
feature J by \cite{RodCom+06}.  It follows the 
most prominent gas tilt to negative latitude at l $>$ 0.  Further
mimicing the larger dust lane, there also appears to be a localized
broad line  over the interval down to 0-velocity at l=3\degr.   

The span of L3 appears contained between the l=5.5\degr\ feature at one
end  (l=5\degr, v = 0 \kms) and the strong-lined gas stretching back 
to Sgr A at the other  (l=1.5\degr, v = 140 \kms).  The region at l=1.3\degr\ 
is well-known for hosting a copious amount of gas and has been suggested
as the contact point for  inflowing material \citep{Fux99}.  
It was recently mapped by \cite{KuhTan+07} and attributed to a 
nascent superbubble, although there is no evidence for recent or
current star formation activity.  A broad-lined, vertically-extended
dustlane standing shock feature forms the core of the gas concentration
at this longitude \citep{Lis06Shred}.

At positive latitude b= 0.5\degr-1.0\degr\ in the right panel, Fig. 2b is
 much leaner and dominated by several (3-4) of the localized broad-lined 
features.  At l$>$ 0\degr, all of the  behavior in Fig. 2b is spectacularly 
contrary to the overall general tilt of the gas layer.
The locus L2, which is not readily apparent in Fig. 1, is 
clearly associated with the b-feature (Clump 2), in fact at its northern 
terminus.  A broad-lined dust lane shock feature at l=1.3\degr\ is
more prominent than in Fig. 2a because of its vertical extension.
The gas at l $< -2$\degr, including Clump 1 of \cite{Ban77}, is a composite 
projection of a rear-side dust lane and the base of a far-side spiral
arm, according to \cite{Fux99}.
 
%3
\begin{figure}
\includegraphics[height=11.3cm]{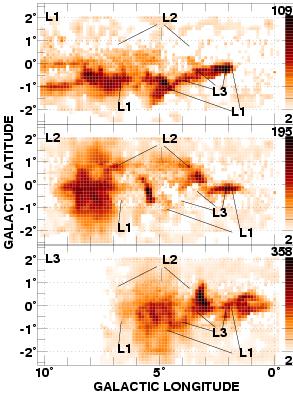}
\caption[]{Maps of CO brightness  in units of K \kms\ 
integrated over the velocities of the
L1, L2 and L3 kinematic dust lane features in Fig. 2.} 
\end{figure} 

\subsection{Locating the dust lane features in space}

To locate the dust lane  features, we created maps of the brightness
integrated across the velocity ranges of their ridges in Fig 2, varying 
the span of the integration with longitude (only).  This is most nearly
definitive for the connecting arm  feature L1 because it occurs
at the extremity of the gas kinematics.  For the other dust lane features, gas 
associated with other aspects of the distribution is unavoidably captured,
but this has the virtue that it exposes the relationships between the
dust lanes and the shredded molecular gas.

Fig 3 has three panels corresponding to integration over the velocity 
spans of the L1-L3 features in Fig. 2.  The L1 panel at top is similar 
to one panel of Fig. 3 of \cite{MarFux+08} because it represents the 
gas in the major dust lane.  At its southernmost terminus at 
l = 5.5\degr, b=-1.8\degr, the gas is  250 pc below the galactic 
equator at a distance of 7.5 kpc  from the Sun
(the dust lane points mostly toward us).

The middle panel integrated over the span of L2 is not particularly
revealing of structure in L2 but it inadvertently captures the 
overall vertical arc of the broad-lined feature at l=5.5\degr\ 
from Fig. 1 and 2a, showing that it terminates below at the locus of 
the L1 dust lane feature \citep{Lis06Shred}.  Some of the bright 
vertically-extended  emission at l = 8\degr-9\degr\ is low-velocity local 
gas.

The lowest panel, representing integration over the L3 feature, shows 
that  Clump 2 at l=3.2\degr\ extends vertically just between L2 and L3.  
This seems unlikely to be an accident, especially considering the 
abruptness with which  L2  terminates in the vicinity of the broad-lined 
feature from Clump 2 in Fig. 2b.   Apparently,
L3 underpins the vertically extended Clump 2 at l = 3.2\degr\  in 
just the same way that L1 underpins the equivalent 
vertically-extended broad-lined feature at l=5.5\degr.

\section{Vertical structure and secular evolution}

In the bar flow picture, the kinematics are spatially and temporally 
sequenced as material streams inward from larger 
radii at higher longitudes. The localized broad-lined features 
generally represent the uptake and shredding of dense material; 
the broad features are seen in the emission of H I,  OH, CO, CS, HCN 
and the many molecules which require higher density for their 
excitation, while the  large scale dust lanes at $|$l$| > 2$\degr\
 are seen 
only in H I and CO (see for instance Fig. 1 of \cite{LeeLee+99}) . 
The multitude of broad-lined features at positive velocity and 
(mostly) positive longitude is related to the presence of a 
stronger or richer dust lane on one side of the Galaxy toward 
the nearer end of the bar.

Presumably there are also a direction and sequence to the vertical 
structure as well.  To interpret the broad-lined feature at l=5.5\degr\ 
it was suggested that material is falling down from regions near the 
equator at the upper extremity of the feature, into the connecting-arm 
dust lane L1 below the plane \citep{Lis06Shred}.  However no source 
of material at the upper extremity was discussed.  By contrast, 
broad-lined Clump 2 at l = 3.2\degr\ has L2 at its upper extremity 
and L3 below, and L2 does not extend inward of l=3\degr.  
This suggests that the source of the gas being shredded at 
Clump 2 is still visible in L2, that the sequence is also
downward and that it is indeed possible for
gas to emerge from a dust lane as well.  Why this should occur
so far from the galactic equator is a mystery but it should
hardly be a surprise if the extremities of Clump 2 and the
feature at l=5.5\degr\ are the site of unusual phenomena.

Understanding the vertical structure of the gas is still a challenge, 
both observationally and theoretically.  Yet, it appears that the
secular evolution of the inner galaxy is actually visible in the
vertical structure and that the tilt of the gas is the result of
this evolution. 

\begin{acknowledgements}

The NRAO is operated by AUI, Inc. under a cooperative agreement 
with the US National Science Foundation.  I thank Tom Dame (CfA) 
for providing the datacube of \cite{BitAlv+97} and thank
him and Pat Thaddeus (CfA) for enabling that work.  Thanks
also to Nemesio Rodriguez-Fernandez and the referee 
for helpful comments.

\end{acknowledgements}

\bibliographystyle{apj}
%\bibliography{mnemonic,shredded}

\end{document}